\documentclass[journal=jacsat,manuscript=article,layout=twocolumn]{achemso}

\newcommand*{\sometext}{We predict the stabilities of $\alpha$-graphynes and their boron nitride analogues($\alpha$-BNyne), which are considered as competitors of graphene and two-dimensional hexagonal BN. Based on first-principles plane wave method, we investigated the stability and structural transformations of these materials at different sizes using phonon dispersion calculations and ab-initio finite temperature, molecular dynamics simulations. Depending on the number of additional atoms in the edges between the corner atoms of the hexagons, $n$, both $\alpha$-graphyne($n$) and $\alpha$-BNyne($n$) are stable for even $n$, but unstable for odd $n$. $\alpha$-graphyne($3$) undergoes a structural transformation, where the symmetry of hexagons is broken. We present the structure optimized cohesive energies, electronic, magnetic and mechanical properties of stable structures. Our calculations reveal the existence of Dirac cones in the electronic structures of $\alpha$-graphynes of all sizes, where the Fermi velocities decrease with increasing $n$. The electronic and magnetic properties of these structures are modified by hydrogenation. A single hydrogen vacancy renders a magnetic moment of one Bohr magneton. We finally present the properties of the bilayer $\alpha$-graphyne and $\alpha$-BNyne structures. We expect that these layered materials can function as frameworks in various chemical and electronic applications.

{\bf Keywords:} graphyne, hydrogenation, phonon modes, molecular dynamics, density functional theory}

\let\oldmaketitle\maketitle
\let\maketitle\relax

\author{V. Ongun \" Oz\c celik}
\affiliation{UNAM-National Nanotechnology Research Center, Bilkent University, 06800 Ankara, Turkey}
\alsoaffiliation{Institute of Materials Science and Nanotechnology, Bilkent University, Ankara 06800, Turkey}
\email{ongunozcelik@bilkent.edu.tr}
\author{S. Ciraci}
\affiliation{UNAM-National Nanotechnology Research Center, Bilkent University, 06800 Ankara, Turkey}
\alsoaffiliation{Institute of Materials Science and Nanotechnology, Bilkent University, Ankara 06800, Turkey}
\alsoaffiliation{Department of Physics, Bilkent University, Ankara 06800, Turkey}
\email{ciraci@fen.bilkent.edu.tr}

\title{Size dependence in the stabilities and electronic properties of $\alpha$-graphyne and its BN analogue}

\begin{document}

\twocolumn[
\begin{@twocolumnfalse}
\oldmaketitle
\begin{abstract}
\sometext
\end{abstract}
\end{@twocolumnfalse}
]

\section{Introduction}

The synthesis of graphene\cite{novoselov2004} has been an important turning point in the study of stable two dimensional (2D) monolayer structures. Graphene has drawn great attention due to its properties such as high chemical stability, mechanical strength and electric conductivity.\cite{geim2007, zhang2005} These unique electronic properties are mainly related to the Dirac cones present in its band structure where the upper and lower cones consisting of the conduction and the valence bands meet at a point on the Fermi level making graphene a semi-metal with zero band gap.

Graphynes, similar to graphene, are two dimensional structures with the inclusion of single and triple bonded carbon atoms between the corner atoms of the honeycomb structure. Much earlier than the synthesis of single layer graphene, Baughman \textit{et al.}\cite{baughman1987} predicted various flakes or molecules of carbon atoms in the graphyne family as layered phases using semi-empirical and empirical atom-atom potential calculations. These finite size nanostructures form from the combination of hexagons with other polygons containing $sp^2$ and $sp$ bonds. Based on first-principles plane wave calculation, Tongay \textit{et al.}\cite{tongay} have predicted various stable 1D, 2D and 3D periodic structures containing carbon atomic chains. Among them, 2D periodic $\alpha$-graphyne have been revealed. Very recently, band structures of the graphyne/graphdyne family with similar behaviors to that of single layer graphene were calculated,\cite{malko2012} showing that neither the existence of hexagonal symmetry nor all atoms being chemically equivalent are prerequisites for the existence of Dirac point in the electronic structures.

Different types of the graphyne family are considered as a new class of 2D materials in the future era of carbon allotropes.\cite{hirsch2010} Finite-size building blocks of these graphyne structures have already been synthesized which is an initial step towards extended structures.\cite{diederich1994, diederich2010, haley2008, kehoe2000, bunz1999, liu2012} Although extended (periodic) 2D structures of $\alpha$-graphyne have not yet been synthesized, the synthesized finite size flakes building blocks are promising for future applications. Theoretical studies and simulations have revealed that different members of the graphyne family can lead to interesting electronic applications and they can be used to construct graphyne-based frameworks and nanotubes. It is also possible to form the single layer, hexagonal boron nitride (h-BN)\cite{bnhoneycomb} analogues of certain types of graphyne, which are called as BNyne throughout this manuscript.\cite{malko2012, narita1998, coluci2003, zhou2011, pan2011} Recently, effects of H, B and N doping into graphyne structures were also investigated.\cite{malko_hetero} Additionally, the physical and chemical properties of graphyne and BNyne families and their future applications can be diversified by varying the sizes of rings and/or by functionalizing them with foreign ad-atoms.

\begin{figure*}
\includegraphics [width=16cm]{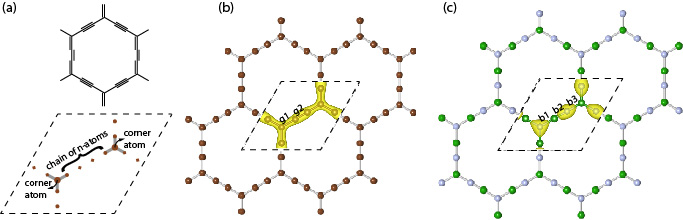}
\caption{Atomic structure of $\alpha$-Graphyne and $\alpha$-BNyne. (a) Schematic representation of $\alpha$-graphyne($2$) and the unit cell used to generate $\alpha$-graphyne($n$), where $n$ is the number of carbon atoms placed between two carbon atoms located at the corners of the hexagon. Two corner atoms of the hexagon have a chain of $n$ atoms between them, such that the unit cell contains $3n+2$ atoms. (b) Atomic structure of single layer, $2D$ $\alpha$-graphyne($2$). The dashed lines delineate the primitive unit cell. The optimized bond lengths are $g1=1.39$\AA~ and $g2=1.23$\AA. The total charge density is shown within the unit cell. Carbon atoms are represented by brown balls. (c) Atomic structure of single layer, $2D$ $\alpha$-BNyne($2$) with blue and green balls representing N and B atoms, respectively. The optimized bond lengths are $b1=1.42$\AA, $b2=1.25$\AA~ and $b3=1.44$\AA. In the charge density plots, the isosurface value is taken as $0.2$ electron/\AA$^3$.}
\label{fig1}
\end{figure*}

Although graphyne allotropes have been widely studied in previous works, their stabilities have been an open question. In this paper we investigate the stabilities of 2D periodic $\alpha$-graphyne structures and their h-BN analogues by using first-principles calculations within density functional theory. An $\alpha$-graphyne structure can be obtained by placing $n$ number of carbon atoms between the corner atoms of the hexagonal graphene structure. We calculate phonon modes and perform ab-initio, finite temperature molecular dynamics(MD) simulations to test the stabilities of different sized $\alpha$-graphynes and $\alpha$-BNynes. We investigate the electronic and mechanical properties of the stable structures. We find that the stabilities, the existence of Dirac points at the Fermi level and the Fermi velocities of massless Fermions depend on the size of the $\alpha$-graphyne, such that $n$=even cases lead to stable structures with graphene like electronic structures, whereas $n$=odd cases lead to instabilities and eventually undergo structural transformations. In addition, the mechanical strengths of $\alpha$-graphynes of various sizes($n=2-4$) were explored and compared with those of other honeycomb like structures. We examine how the electronic properties are affected by hydrogen atoms adsorbed at the corner carbon atoms of hexagons alternatingly from the top and bottom sides and thus formed a structure analogous to graphane.\cite{graphane} In addition, we provide a comprehensive comparison of cohesive energies with those of constituent 1D and 2D allotropes of carbon. We finally investigate the cohesive energies and electronic properties of  bilayer structures of these materials. We also perform similar analysis for the h-BN analogues of graphynes, \textit{i.e.} $\alpha$-BNynes. Our study revealed a new, non-hexagonal $\alpha$-graphyne structure, where the symmetries associated with hexagons are broken.

\section{Method}

In our calculations we used the state-of-the-art first-principles plane-wave calculations within the density-functional theory combined with ab-initio, finite temperature MD calculations using projector augmented wave potentials.\cite{blochl1994} The numerical plane wave calculations were performed using the VASP package.\cite{kresse1993, kresse1996} The exchange correlation potential was approximated by the generalized gradient approximation with the van der Waals correction.\cite{perdew1992, grimme2006} A plane-wave basis set with energy cutoff value of $600 eV$ was used. The Brillouin zone (BZ) was sampled in the \textbf{k-}space within the Monkhorst-Pack scheme,\cite{monkhorst1976} and the convergence of the total energies and magnetic moments with respect to the number of k-points were tested. In the self consistent total energy calculations and the MD simulations, the BZ was sampled by ($17 \times 17 \times 1$) mesh points in the \textbf{k-} space. The convergence criterion for energy was chosen as $10^{-5}eV$ between two consecutive steps. In the geometry relaxation and band structure calculations, the smearing value for all structures was taken as $0.01eV$. The pressure on each system was kept smaller than $\sim2kBar$ per unit cell in the calculations. In the ab-initio MD calculations, the time step was taken as $2.5fs$ and atomic velocities were renormalized to the temperature set at $T=500K$ and $T=1000K$ at every $40$ time steps. In the MD stability tests, the simulations were run for $5ps$. The phonon dispersion curves were calculated using plane wave methods as implemented in the PWSCF package.\cite{giannozzi2009}

\section{$\alpha$-Graphyne and $\alpha$-BNyne}

\subsection{Structure}

\begin{figure}
\includegraphics [width=8.5cm]{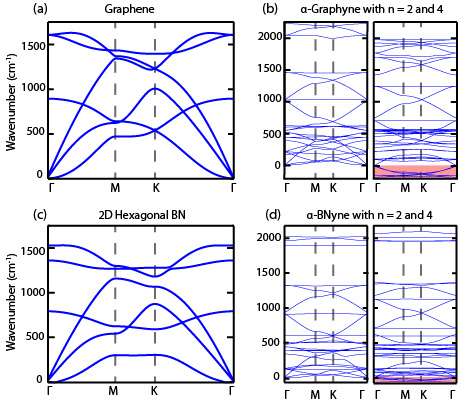}
\caption{Calculated phonon bands. (a) Graphene. (b) $\alpha$-Graphyne with $n=$2 and $n=$4. (c) Single layer h-BN. (d) $\alpha$-BNyne with $n=2$ and $n=$4. The dispersion curves for $n$=2 have totally positive phonon modes which is an indication of their stability. On the other hand, $n$=4 cases have modes with imaginary frequencies, which are marked with the shaded regions and will be discussed in the text. Phonon bands of unstable structures, such as $n=1$ and $n=3$ are not shown.}
\label{fig2}
\end{figure}

\begin{figure*}
\includegraphics [width=16cm]{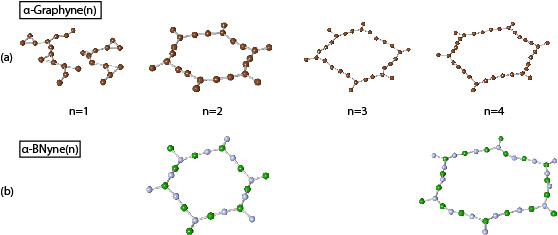}
\caption{Snapshots of the MD simulations performed for $5ps$ at $T=1000K$. (a) $\alpha$-graphyne($n$) $n=$1-4 structures. The structures are stable for $n=$2 and $n=$4, although buckled in the vertical plane. On the other hand, $n=1$ case breaks into carbon atomic strings, and hence are totally unstable. $\alpha$-graphyne($3$) undergoes a structural transformation, whereby it acquires stability by changing the number of C atoms to $n=2$, $n=3$ and $n=4$ in the adjacent edges of hexagon. (b) Sections of $\alpha$-BNyne($n$). Both $n=$2 and $n=$4 cases remain stable during MD simulations. $n=$1 and $n=$3 cases are missed, since $\alpha$-BNyne($n$) cannot be formed with odd $n$. Note that only a single ring from the periodic structures are shown here.}
\label{fig3}
\end{figure*}

\begin{figure}
\includegraphics [width=7cm]{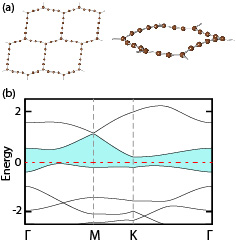}
\caption{(a) Top and side views of the optimized geometry of the structurally transformed, stable $\alpha$-graphyne($3$). The planar, unstable $\alpha$-graphyne($3$) undergoes a structural transformation and acquires stability by breaking the symmetry of hexagons and eventually by buckling. The adjacent edges of the hexagon has $n$=2, 3 and 4 carbon atoms, respectively. (b) Electronic structures of the buckled $\alpha$-graphyne($3$). The buckling leads to a gap opening of 0.42eV.}
\label{fig4}
\end{figure}

We begin our analysis by investigating the structural and geometrical properties of $\alpha$-graphynes at different ($n$) sizes. As shown in \ref{fig1}(a), $\alpha$-graphyne($n$) structure has a hexagonal unit cell like graphene, but there are $3n+2$ carbon atoms in its unitcell instead of $2$. In other words, we have the normal graphene-like structure with chains\cite{carbonchain1} consisting of $n$ carbon atoms between the corner atoms of the hexagon on each side. Contrary to graphene, all of the C-C bonds are not equivalent to each other in these structures, but there exists bonds with different lengths and charge densities as shown in \ref{fig1}(b). Variances of bonds originate from different types of bonding, $sp^2$ and $sp^1$ (+$\pi$) between carbon atoms. Therefore, the carbon atoms are no longer chemically equivalent to each other as they were in the case of graphene.

The h-BN analogue of this structure, namely $\alpha$-BNyne, can be easily obtained by replacing the carbon atoms with B and N consecutively. The electronegativity numbers of B and N atoms are 2.0 and 3.0, respectively according to Pauling's electronegativity scale.\cite{pauling} Hence the electronic charge is transferred from B to adjacent N in the ionic bonds of $\alpha$-BNyne. Here, there are three different types of bonds; two at the corners of the hexagon, one at the center of edges. For B-N bonds at the corners, one is the case where N is at the corner of the hexagon and B in the edge as denoted by $b1$ in \ref{fig1}(c), and the other is the opposite of this and denoted by $b3$. Note that, if we want to preserve the 2D-BN hexagonal structure  such that the corner atoms of the hexagons are consecutively B and N, we must have even number of atoms between the corner atoms. That is, we can only have $n=$even scenarios for the $\alpha$-BNyne($n$) cases. Odd $n$ values would either lead to same type of atoms on the corners of the hexagons or two identical atoms(B-B or N-N) next to each other. For this reason, we restrict ourselves with even $n$ values for $\alpha$-BNyne($n$). The periodic geometries of these structures, and the unit cells used to generate them are shown in \ref{fig1}(b) and \ref{fig1}(c) for $\alpha$-graphyne(2) and $\alpha$-BNyne(2) cases, respectively.

\subsection{Stability}

Up to now, finite building blocks of graphyne allotropes\cite{diederich1994, diederich2010, haley2008, kehoe2000, bunz1999, liu2012} and also the segments of atomic chains consisting of $n$ carbon atoms\cite{eisler, meyer, chalifoux} have been synthesized, which may be taken as an indication of the stabilities of their extended structures. Here, the main issue is whether 2D single layer, periodic structure of $\alpha$-graphyne is stable or not.
As an initial step we calculated the cohesive energies of these structures, which is defined as the energy required to form separated neutral atoms in their ground electronic state from the condensed state at $0K$ and $1atm$.\cite{kittel} Cohesive energies per atom of the carbon allotropes (namely graphene, $\alpha$-graphyne($n$) and cumulene or infinite carbon atomic chain\cite{poly}) are obtained using the expression

\begin{equation}
E_{C}=\{ pE_{T}[C]-E_{T}[C_{allotrope}] \} / p,
\end{equation}

in terms of the optimized total energies of single carbon atom in its magnetic ground state $E_{T}[C]$, and of the related carbon allotrope $E_{T}[C_{allotrope}]$, with $p$ being the number of carbon atoms in the unit cell. In the case of BN structures, the cohesive energy is calculated for per B-N pair using

\begin{equation}
E_{C} = \{ qE_{T}[B]+q E_{T}[N]-E_{T}[BN_{allotrope}] \} / q
\end{equation}

in terms of optimized total energies of single B atom $E_{T}[B]$, N atom $E_{T}[N]$ and of the related BN allotrope, with $q$ being the number of B-N pairs in the unit cell. Here (BN)$_{allotrope}$ stands for one of the structures, such as BN chain,\cite{bnchain} h-BN and $\alpha$-BNynes. The calculated cohesive energies of graphene, $\alpha$-graphyne(n), cumulene and their h-BN analogues are given in Table 1. We note that the cohesive energies of $\alpha$-graphyne($n$) and $\alpha$-BNyne($n$) are smaller than their parent, single layer graphene and h-BN honeycomb structures, respectively. Also expectantly, the cohesive energies of $\alpha$-graphyne($n$) and -BNyne($n$) decrease with increasing $n$, since the cohesive energy of the linear atomic chain (either cumulene or BN) is smaller than the corresponding 2D single layer, honeycomb structure (respectively, either graphene or h-BN). Like graphene and carbon atomic chains, the calculated cohesive energies suggest that $\alpha$-graphyne structures correspond to local minima on the Born-Oppenheimer surface. Accordingly, even the network of carbon atomic chains consisting of diverse number of atoms $n$, which are connected by three folded $sp^2$-orbitals at the nodes can be in local minima.\cite{carbonchain} In principle, these networks can form either multiply connected polygons(rings) or tree-like structure. Here we note that the cohesive energies of $\alpha$-graphyne($n$) for $n=$1 and $n=$3 calculated by restricting the structure in a plane are found smaller than that of $n=$2 and $n=$4. As clarified in the following analysis of stability using phonon calculations, $n=$odd number structures are unstable despite their positive cohesive energy. Here, smaller $E_{C}$ for $n=$1 and $n=$3 relative to that of $n=$2 and $n=$4 indicate instability, but other analysis like calculation of phonon dispersion and MD simulations are needed for a reliable test of stability.

\begin{table}
\caption{Cohesive energies($E_{C}$) of optimized graphene, $\alpha$-graphyne($n$) $n$=1-4, cumulene structures and their h-BN analogues. The cohesive energies of the BN structures are given per number of B-N pairs.}
\label{table: 1}
\begin{center}
\begin{tabular}{lc}
\hline  \hline
Structure & $E_{C}$ (eV) \\
\hline
\hline
Graphene & 8.11 \\
Carbon chain & 7.10 \\
$\alpha$-graphyne($1$) & 6.21 \\
$\alpha$-graphyne($2$) & 7.35 \\
$\alpha$-graphyne($3$) & 6.82 \\
$\alpha$-graphyne($4$)& 7.12 \\
2D BN & 14.52 \\
BN chain & 12.80 \\
$\alpha$-BNyne($2$) & 13.26 \\
$\alpha$-BNyne($4$) & 12.90 \\

\hline
\hline
\end{tabular}
\end{center}
\end{table}

A state-of the art calculations of phonon frequencies of $\alpha$-graphynes and $\alpha$-BNynes are carried out for all modes as a function of the \textbf{k}-points in the BZ. When the calculated frequencies of all the modes are positive, the structure is identified to be stable. We note that the constituent allotropes, namely graphene, h-BN, carbon and BN chains, have positive frequencies for all modes in their BZ, and hence they are stable. As seen in \ref{fig2}, all calculated frequencies of $\alpha$-graphyne($2$) and $\alpha$-BNyne($2$) are positive. Moreover, due to the segments of atomic carbon and BN chains in $\alpha$-graphyne($2$) and $\alpha$-BNyne($2$), both structures have phonon modes with frequencies higher than those of single layer graphene and h-BN. As a matter of fact, the maximum frequency of the longitudinal optic mode (LO) of carbon chain\cite{carbonchain} can be as high as $\sim$ 2400 cm$^{-1}$.

The situation is seemingly different for $\alpha$-graphyne($4$) and $\alpha$-BNyne($4$), since some of the acoustic modes have imaginary frequencies. These modes shaded out in \ref{fig2}(b) and (d) correspond to imaginary frequencies and normally would indicate instabilities. However, the imaginary frequencies may arise also as an artifact of numerical calculations. To obtain the real frequencies of soft modes in an open structure comprising many atoms, one need to perform calculations by taking into account the distant neighbors and use very high numerical accuracies. Thus, we believe that the calculated imaginary frequencies of soft modes in this case are artifacts of numerical phonon calculations since the segments of chains consisting of even carbon atoms have been found to be stable\cite{carbonchain} and they become even more stabilized if they are connected to carbon atoms with three folded $sp^2$ orbitals. Also, it should be noted that all $\alpha$-graphyne($n$) and $\alpha$-BNyne($n$) with large $n$ can be stabilized at finite-size, since long wavelength acoustical modes are discarded.

In order to further investigate the issue of stability, we performed MD simulations at $T=1000K$ for $5ps$. The atomic structures obtained after the MD calculations are presented in \ref{fig3}. It turns out that for $n=$2 and $n=$4, $\alpha$-graphyne($n$) and $\alpha$-BNyne($n$) remain stable after $5 ps$ of MD simulation even at high temperatures. Even though $5 ps$ is a short time interval, it is long enough for ab-initio calculations at the temperature as high as $T=1000K$ to provide evidence for stability. The analysis of atomic structures also suggests that $\alpha$-graphyne($4$) and $\alpha$-BNyne($4$) prefer to buckle in the vertical direction. Also in some cases, when specific structures are unstable in planar geometry they might undergo structural transformations to attain stability. Silicene (\textit{i.e.} Si in graphene structure) is a crucial example for stability gained by buckling.\cite{silicene} Here we assure that the buckling occurs due high temperature for the following reasons: (i) The optimization of buckled structure at T=0K ended with planar structure; (ii) similar buckling effects were also observed at segments of carbon atomic chains self-assembled on graphene and h-BN which bowed at high temperatures.\cite{can_chain, ongun_chain, growth}

It was found that $\alpha$-graphyne($1$) is totally unstable and is dissociated into carbon atomic strings. $\alpha$-graphyne($3$) presents an interesting situation; it undergoes a structural transformation and acquires stability by changing the number of carbon atoms to $n=$2, $n=$3 and $n=$4 in the adjacent edges of hexagon as shown in \ref{fig4}. This structural transformation is derived to maintain the proper bond order of finite size carbon chains constituting the edges of hexagon.\cite{carbonchain} While the carbon atoms at the corners are always forming single bonds with the adjacent carbon atoms, the second atom from the corner by itself has to make a triple bond with the adjacent carbon atom, which is at the other side in the edge of the hexagon. At the end, the correct bond order of carbon atoms are preserved. However, this structural transformation, which modifies the geometric structure and breaks the symmetry of hexagons, reveals a new buckled allotrope carbon atom in 2D with a band gap of 0.42eV.

\begin{figure}
\includegraphics [width=8cm]{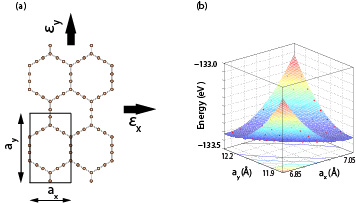}
\caption{(a)$\alpha-$Graphyne($2$) structure in rectangular unit cell with its lattice constants $a_x$ and $a_y$. $\epsilon_x$ and $\epsilon_y$ are the strains in $x$ and $y$ directions, respectively. (b) 3D plot of the energy values corresponding to different $a_x$ and $a_y$ values.}
\label{fig5}
\end{figure}

\begin{figure*}
\includegraphics [width=16cm]{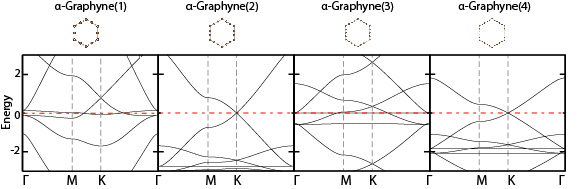}
\caption{Electronic band structures of $\alpha$-graphyne($n$) for $n=1,2,3$ and $4$. All of the band structures contain Dirac points, while they are shifted above the Fermi level for $n=1$ and $3$. $n=1$ and $3$ cases also have Dirac points away from the high symmetry $K-$point. The zero of energy is set to the Fermi level. Note that, the electronic structures of the $n=1$ and $n=3$ structures are presented for the sake of completeness although they are unstable in the planar configuration, but $n=3$ structure acquires stability in the buckled geometry as discussed in the text.}
\label{fig6}
\end{figure*}

\begin{figure}
\includegraphics [width=8.5cm]{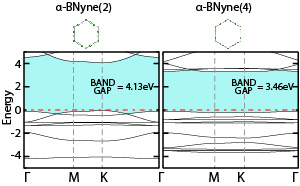}
\caption{Electronic band structures of stable $\alpha$-BNyne($n$) for $n=2$ and $4$. Note that as $n$ increases, the band gap decreases. The maximum energy of the valence band is set to zero.}
\label{fig7}
\end{figure}

\subsection{Mechanical properties}

Having found the stable $\alpha$-graphyne and $\alpha$-BNyne structures, we next calculate their mechanical strengths. A common way of expressing the mechanical properties of two dimensional materials is to calculate their in-plane stiffness, Poisson's ratio and Young's modulus values. For this purpose, we used a rectangular unit cell in the $xy$-plane and applied tension in both $x$ and $y$ directions, as shown in \ref{fig5}(a). We varied the lattice constants, $a_x$ and $a_y$, between $\pm 0.03\%$ of the optimized values and calculated the energy value for each grid point obtained. In the end, we have obtained energy values for $225$ grid points, which are plotted in \ref{fig5}(b) for $\alpha$-graphyne($2$). The strain energy $E_{s}$, at each point is calculated by subtracting the total energy at that point from the equilibrium total energy.  It has been shown that this method of calculating the strain energy provides reliable predictions for graphene, and 6-6-12 graphyne.\cite{topsakal2010, cranford2011, kang2011}

The in-plane stiffness, which is commonly used measure of strength for 2D materials can be expressed as

\begin{equation}
 C=\frac{1}{A_o} \times \frac{\partial^2E_s}{\partial \epsilon^2}
\end{equation}

where $E_s$ is the strain energy, $A_o$ is the equilibrium area and $\epsilon$ is the axial strain calculated by $\Delta a_{x,y}/a_{x,y}$, $a$ being the lattice constant in the $x$ or $y$ direction.

The in-plane stiffness values of  $\alpha$-graphyne($2$), $\alpha$-graphyne($4$), $\alpha$-BNyne($2$) and $\alpha$-BNyne($4$) were calculated as $21 N/m$, $16 N/m$, $19 N/m$, and $14 N/m$, respectively. These values are much lower than the in-plane stiffness of graphene, which is $\sim340N/m$.\cite{topsakal2010, lee2008} As seen from these results, the implementation of new atoms between atoms at the corners of hexagons decreases the mechanical strength of graphene dramatically, which is a direct consequence of the decrease in the average  coordination number of carbon atoms. The Poisson's ratio, which is defined as the ratio of the transverse strain to the axial strain, $\nu=-\epsilon_{trans} / \epsilon_{axial}$, was calculated as $0.88$, $0.86$, $0.89$ and $0.85$ for $\alpha$-graphyne(2), $\alpha$-graphyne(4), $\alpha$-BNyne(2) and $\alpha$-BNyne(4), respectively. By assuming an equivalent thickness with graphene, the Young's modulus values were calculated, respectively as $61GPa$, $48GPa$, $52GPa$ and $42GPa$ for these structures. We also note that in-plane stiffness, Poisson's ratio and Young's modulus values of $\alpha$-graphyne($n$) and $\alpha$-BNyne($n$) decrease with increasing $n$. Also, the values calculated for $\alpha$-BNyne($n$) are lower than those calculated for $\alpha$-graphyne for each case with equivalent $n$.

\subsection{Electronic structure}

Earlier, Tongay \textit{et al.}\cite{tongay} showed that the electronic structure of $\alpha$-graphyne($2$) with two bands crossing the Fermi level at $K$- and $K^\prime$-points of the BZ is similar to that of graphene. The presence of Dirac cones in the band structure of $\alpha$-graphynes, as well as $\beta-$graphyne and 6-6-12 graphyne have also been recently reported,\cite{malko2012} and it was demonstrated that the existence of Dirac points and cones is not a unique property of graphene. The crossing of bands at the Fermi level and the formation of Dirac cones were also investigated for other structures.\cite{sahin2011} It was pointed out that bare structures, large defects and ad-atoms on graphene can have Dirac cones if their periodic patterns comply with a specific symmetry. Here, we present the electronic energy  band structure of $\alpha-$graphyne($n$) for $n=$1,2,3 and 4 in \ref{fig6}. The existence of Dirac points is also seen here for all of these graphynes. Note that, the Dirac point lies at the Fermi level for the stable $n$=2 and $n$=4 structures, whereas it is shifted above from the Fermi level for the unstable $n$=1 and $n$=3. We note that high electron density at the Fermi level of $\alpha-$graphyne($n$) for $n$=1 and $n$=3 can be attributed to their instability. The electronic energy near the K-point of the BZ is linear with respect to \textbf{q}=\textbf{K}-\textbf{k}, which leads

\begin{equation}
 E(\pm) = \pm \hbar v_F \textbf{q} + O[(\textbf{q}/\textbf{k})^{2}],
\end{equation}

where $v_F$ is the Fermi velocity and \textbf{K} is the wave vector corresponding to $K$- and $K^\prime$-points of the BZ. Then, for the stable $\alpha$-graphyne($2$) and $\alpha$-graphyne($4$) structures, the first derivatives of their $\pi$ bands near the $K$-point of the BZ were calculated as 29.4 $eV$\AA~and 23.4 $eV$\AA~, which have the same order of magnitude obtained for graphene, 34.6 $eV$\AA~. We also estimate the Fermi velocities as $v_{F} \sim 8.3 \times 10^5$ m/s for graphene, $v_{F} \sim 7.1 \times 10^5$ m/s for $\alpha$-graphyne($2$) and $v_{F} \sim 5.6 \times 10^5$ m/s for $\alpha$-graphyne($4$). Accordingly, the Fermi velocities at $K$- and $K^{'}$-points decrease with increasing $n$. Noting the electron-hole symmetry of Eq(4) near $K$-point, $\alpha$-graphyne($2$) and $\alpha$-graphyne($4$) are found to be ambipolar. As discussed in Sec. III B, $\alpha$-graphyne($3$) having metallic state undergo structural transformation where hexagons forming a honeycomb structure transforms to rectangle like structures as shown in the third panel of \ref{fig3}. When planar, this structure attains high density of states at the Fermi level. However, through the buckling of the atomic structure it undergoes a metal-insulator transition by opening an indirect band gap of 0.42 eV (See ~\ref{fig4}).

On the other hand, reminiscent of the electronic structure of 2D single layer h-BN, the band structures of $\alpha$-BNyne($2$) and  $\alpha$-BNyne($4$) have wide band gaps as shown in \ref{fig7}. It can be seen that the band gap decreases with increasing values of $n$. The lowest conduction band and the highest valence band states originate from B-$p_{z}$ and N-$p_{z}$ orbitals.

\section{Hydrogenation}

\begin{figure}
\includegraphics [width=8cm]{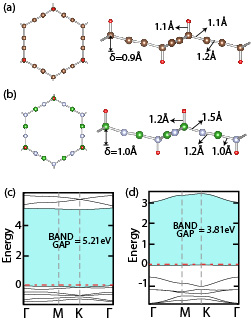}
\caption{(a-b) Hydrogenated $\alpha$-graphyne($2$) and $\alpha$-BNyne($2$). Top and side views of the optimized hydrogenated structures are shown by the ball and stick models, where C, N, B and H atoms are represented by brown, blue, green and red balls, respectively. (c-d) Electronic structures of hydrogenated $\alpha$-graphyne($2$) and $\alpha$-BNyne($2$). Note that, in contrast to the semimetalic $\alpha$-graphyne($2$), the hydrogenated structure has a wide band gap of 5.2eV. The maximum energy of the valence band is set to zero.}
\label{fig8}
\end{figure}

As an immediate application $\alpha$-graphyne($n$) and $\alpha$-BNyne($n$), one may consider their chemical conversion through the coverage of H (hydrogenation), F (fluorination), Cl (chlorination). Here we present our results regarding the hydrogenation of $\alpha$-graphyne($2$) and $\alpha$-BNyne($2$). The hydrogenation of graphene which produces graphane, and the dehydrogeneation of graphane are already well known process\cite{sofo2007, graphane}. Subsequently, hydrogenation of single layer BN was also theoretically studied which lead into to BN-phane.\cite{averill2009}  In a similar way, we attach single hydrogen atoms to the corner carbon atoms of $\alpha$-graphyne($2$) and B and N atoms at the alternating corners of $\alpha$-BNyne($2$) from top and bottom alternatingly, as illustrated in \ref{fig8}(a-b). Upon hydrogenation, both $\alpha$-graphyne($2$) and $\alpha$-BNyne($2$) structures relax into a buckled geometry with buckling distances of 0.92 \AA~ and 1.03 \AA. These buckled geometries are similar to the buckled geometries of graphane and BN-phane. However, the buckling increases from 0.45 \AA~ to 0.92 \AA~ as we go from graphane to hydrogenated $\alpha$-graphyne($2$), as a result of the increased length of the edges of hexagons.  Hydrogenation also alters the electronic structures of these systems, as presented in \ref{fig8}(c-d). Consequently, the semimetalic $\alpha$-graphyne($2$) structure attains a wide band gap of 5.2 eV upon hydrogenation. While hydrogenated $\alpha$-graphyne($2$) has nonmagnetic ground state, a single hydrogen vacancy renders a magnetic moment of 1 $\mu_{B}$

Hydrogenated $\alpha$-BNyne($2$) also has a wide band gap of 3.3eV. The band gap opening effect of hydrogenation on monolayer structures is in accordance with prior calculations on graphane\cite{sofo2007} and hydrogenated BN,\cite{averill2009} which have band gaps of 3.5eV and 4.8eV, respectively.

\section{Bilayer Structures}

\begin{figure}
\includegraphics [width=8.5cm]{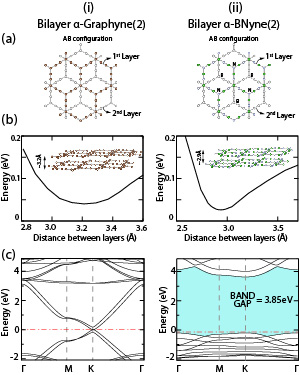}
\caption{Bilayer $\alpha$-graphyne($2$) and its BN analogue bilayer $\alpha$-BNyne($2$) are shown in columns $i$ and $ii$, respectively. (a) Top view of the optimized two layer structures. Both bilayer $\alpha$-graphyne and $\alpha$-BNyne have AB type of stacking geometry, which is more favorable than the AA stacking. In the ball and stick model C, B and N atoms, respectively are represented by brown, green and blue balls and the all of the atoms in the bottom layer are shown in gray balls. (b) Variation of energy as a function of the layer-layer distance. (c) Electronic band structures of $\alpha$-graphyne($2$) and $\alpha$-BNyne($2$).}
\label{fig9}
\end{figure}

Here we address the question of whether $\alpha$-graphyne and $\alpha$-BNyne can form layered structures similar to graphite and hexagonal BN, or not. We place their bilayers as shown in figure \ref{fig9} and explore equilibrium geometries, binding energies and electronic structures. We begin the analysis by placing two single layers of $\alpha$-graphyne(2) sheets on top of each other with AA stacking (\textit{i.e.}  hexagons in both layers face each other) and AB stacking (\textit{i.e.} first layer is shifted laterally to the centers of hexagons in the second layer) geometries. We alter the interlayer distances until we achieve the minimum energy values. The calculated minimum energy geometries indicate that AB type of stacking is more favorable than AA type by $46meV$. This is a behavior similar to bilayer graphene structure or graphite. The optimized interlayer distance is calculated as $3.12$\AA, which is less than the interlayer distance of graphite, $~3.35$\AA. This is mainly related to the less dense arrangement of carbon atoms on the graphyne surface as compared to the graphene, which results in lower surface energy and hence closer equilibrium distance. Similar results were also found in a previous interlayer distance and stacking studies made on different graphyne allotropes.\cite{cranford2011, zheng2012}

A convenient procedure for calculating the interlayer binding energy for layered structures is subtracting the minimum energy of the bilayer structure from the sum of energies of separated individual layers.\cite{bjorkman2012} Applying this method, we calculate the binding energy of bilayer $\alpha$-graphyne(2) as $220meV (27.5meV/atom)$. We repeat the same analysis for bilayer $\alpha$-BNyne(2), which also has an AB type double layer geometry as shown in the second column  of \ref{fig9}(a). For this case, the energy difference between AB stacking and the AA stacking is in favor of AB by $70meV$, the interlayer distance is $2.9$\AA~ and the binding energy is $128meV (16meV/atom)$. The calculated binding energies of bilayer $\alpha$-graphyne(2) and bilayer $\alpha$-BNyne(2) are in the range of the calculated binding energies of bilayer graphene $(48meV/atom)$ and bilayer h-BN $(31.5meV/atom)$, respectively. The variations of total energies around the minimum energy values as a function of the interlayer distances are shown in \ref{fig9}(b). Like graphite and layered BN, small interlayer binding energies consist of mainly from the van der Waals interaction with even smaller chemical interaction component. Nevertheless, such a weak binding is enough to maintain the bilayer structure at low temperature. Additionally, 3D layered structures of $\alpha$-graphyne and $\alpha$-BNyne can form at room temperature. These arguments are corroborated by the MD simulations performed at different temperatures showing that the bilayer structures form stable interlayer binding near room temperature, but the layers move away from each other at high temperatures ($T=500K$  and $1000K$). These results imply that multilayers of $\alpha$-graphyne and $\alpha$-BNyne, even their 3D layered structures can form.

We finally calculate the electronic structures of these double layered structures. The band structure of bilayer $\alpha$-graphyne($2$) is shown in \ref{fig9}(c). Note that, the bands are split due to the couplings between layers and the numbers of energy bands are doubled as compared to the single $\alpha$-graphyne($2$) sheet. Notably, the bands are no longer linear around the $K$-point, but parabolic. This kind of structure is reminiscent of the electronic structure of bilayer graphene or graphite.\cite{partoens2006} As opposed to single layer, the bands no longer touch at the K point but there is a small band gap of $10meV$. The electronic structure for the bilayer $\alpha$-BNyne($2$) is also presented in \ref{fig9}(c). Similar to the single layer, the double layered structure also has a wide band gap.

\section{Conclusion}

In conclusion, we investigated the cohesion, structural stabilities, electronic structures and mechanical properties and functionalization by adatoms of $\alpha$-graphyne($n$) structures along with their h-BN analogues $\alpha$-BNyne($n$). By performing both phonon frequency and finite temperature molecular dynamics analysis for $n=1-4$ cases, we showed that both $\alpha$-graphyne and $\alpha$-BNyne are stable structures for even $n$, but unstable for odd $n$. Interestingly, $\alpha$-graphyne($3$) undergoes a structural transformation to acquire stability, whereby the symmetry of hexagons forming honeycomb structure is broken. Thus, our study clarified the question whether 2D periodic $\alpha$-graphyne  and $\alpha$-BNyne structure can be stable or not. We also calculated the electronic structures for each of these materials and showed that $\alpha$-graphyne($n$) structures having Dirac cones are ambipolar and their Fermi velocities decrease with increasing $n$. It is also implied that, all atoms being chemically equivalent is not a prerequisite for the existence of Dirac cones in the electronic structure. Upon hydrogenation, the Dirac cones are replaced by a large band gap. Since the formation of a single hydrogen vacancy renders a magnetic moment of 1 $\mu_{B}$, magnetic nanomaterials can be designed by creation of domains of hydrogen vacancies. Our calculations of mechanical properties revealed $\alpha$-graphyne and $\alpha$-BNyne are not as stiff as graphene and the single layer h-BN, but they are strong enough to sustain the technological applications.

We finally showed that it is also possible to have double layers of $\alpha$-graphyne and $\alpha$-BNyne structures. Both of these bilayer structures have AB type of stacking. It was found that the electronic structure of bilayer $\alpha$-graphyne has a gap opening of $10 meV$ at the $K-$point of the BZ as opposed to single layer $\alpha$-graphyne(2). On the other hand, the electronic structure of bilayer $\alpha$-BNyne(2) contains a wide band gap, similar to single layer BN structure.

Briefly, we demonstrated that $\alpha$-graphyne and $\alpha$-BNyne nanostructures can form stable and durable 2D extended structures with interesting chemical and physical properties, which are scaled by $n$. We believe that, these stable 2D carbon and BN allotropes will attract interest because of their unique properties in the near future. In particular, they can be utilized as structural frameworks for various spintronic\cite{engin2006} and chemical applications.

\section{Acknowledgements}

The computational resources have been provided by TUBITAK ULAKBIM, High Performance and Grid Computing Center (TR-Grid e-Infrastructure) and UYBHM at Istanbul Technical University through Grant No. 2-024-2007. This work was supported by the Academy of Sciences of Turkey(TUBA). The authors thank S. Cahangirov and M. Topsakal for comments and discussions.

\bibliography{graphyne_JPCC.bbl}

\providecommand*{\mcitethebibliography}{\thebibliography}
\csname @ifundefined\endcsname{endmcitethebibliography}
{\let\endmcitethebibliography\endthebibliography}{}
\begin{mcitethebibliography}{52}
\providecommand*{\natexlab}[1]{#1}
\providecommand*{\mciteSetBstSublistMode}[1]{}
\providecommand*{\mciteSetBstMaxWidthForm}[2]{}
\providecommand*{\mciteBstWouldAddEndPuncttrue}
  {\def\EndOfBibitem{\unskip.}}
\providecommand*{\mciteBstWouldAddEndPunctfalse}
  {\let\EndOfBibitem\relax}
\providecommand*{\mciteSetBstMidEndSepPunct}[3]{}
\providecommand*{\mciteSetBstSublistLabelBeginEnd}[3]{}
\providecommand*{\EndOfBibitem}{}
\mciteSetBstSublistMode{f}
\mciteSetBstMaxWidthForm{subitem}{(\alph{mcitesubitemcount})}
\mciteSetBstSublistLabelBeginEnd{\mcitemaxwidthsubitemform\space}
{\relax}{\relax}

\bibitem[Novoselov et~al.(2004)Novoselov, Geim, Morozov, Jiang, Zhang, Dubonos,
  Grigorieva, and Firsov]{novoselov2004}
Novoselov,~K.~S.; Geim,~A.~K.; Morozov,~S.~V.; Jiang,~D.; Zhang,~Y.;
  Dubonos,~S.~V.; Grigorieva,~I.~V.; Firsov,~A.~A. \emph{Science}
  \textbf{2004}, \emph{306}, 666--669\relax
\mciteBstWouldAddEndPuncttrue
\mciteSetBstMidEndSepPunct{\mcitedefaultmidpunct}
{\mcitedefaultendpunct}{\mcitedefaultseppunct}\relax
\EndOfBibitem
\bibitem[Geim and Novoselov(2007)]{geim2007}
Geim,~A.~K.; Novoselov,~K.~S. \emph{Nat. Mater.} \textbf{2007}, \emph{6},
  183--191\relax
\mciteBstWouldAddEndPuncttrue
\mciteSetBstMidEndSepPunct{\mcitedefaultmidpunct}
{\mcitedefaultendpunct}{\mcitedefaultseppunct}\relax
\EndOfBibitem
\bibitem[Zhang et~al.(2005)Zhang, Tan, Stormer, and Kim]{zhang2005}
Zhang,~Y.; Tan,~Y.~W.; Stormer,~H.~L.; Kim,~P. \emph{Nature (London)}
  \textbf{2005}, \emph{438}, 201--204\relax
\mciteBstWouldAddEndPuncttrue
\mciteSetBstMidEndSepPunct{\mcitedefaultmidpunct}
{\mcitedefaultendpunct}{\mcitedefaultseppunct}\relax
\EndOfBibitem
\bibitem[Baughman et~al.(1987)Baughman, Eckhardt, and Kertesz]{baughman1987}
Baughman,~R.~H.; Eckhardt,~H.; Kertesz,~M.~J. \emph{J. Chem. Phys.}
  \textbf{1987}, \emph{87}, 6687--6698\relax
\mciteBstWouldAddEndPuncttrue
\mciteSetBstMidEndSepPunct{\mcitedefaultmidpunct}
{\mcitedefaultendpunct}{\mcitedefaultseppunct}\relax
\EndOfBibitem
\bibitem[Tongay et~al.(2005)Tongay, Dag, Durgun, Senger, and Ciraci]{tongay}
Tongay,~S.; Dag,~S.; Durgun,~E.; Senger,~R.~T.; Ciraci,~S. \emph{J. Phys. :
  Condens. Matter} \textbf{2005}, \emph{17}, 3823--3836\relax
\mciteBstWouldAddEndPuncttrue
\mciteSetBstMidEndSepPunct{\mcitedefaultmidpunct}
{\mcitedefaultendpunct}{\mcitedefaultseppunct}\relax
\EndOfBibitem
\bibitem[Malko et~al.(2012)Malko, Neiss, Vines, and Gorling]{malko2012}
Malko,~D.; Neiss,~C.; Vines,~F.; Gorling,~A. \emph{Phys. Rev. Lett.}
  \textbf{2012}, \emph{108}, 086804\relax
\mciteBstWouldAddEndPuncttrue
\mciteSetBstMidEndSepPunct{\mcitedefaultmidpunct}
{\mcitedefaultendpunct}{\mcitedefaultseppunct}\relax
\EndOfBibitem
\bibitem[Hirsch(2010)]{hirsch2010}
Hirsch,~A. \emph{Nature Mater.} \textbf{2010}, \emph{9}, 868--871\relax
\mciteBstWouldAddEndPuncttrue
\mciteSetBstMidEndSepPunct{\mcitedefaultmidpunct}
{\mcitedefaultendpunct}{\mcitedefaultseppunct}\relax
\EndOfBibitem
\bibitem[Diederich(1994)]{diederich1994}
Diederich,~F. \emph{Nature (London)} \textbf{1994}, \emph{369}, 199--207\relax
\mciteBstWouldAddEndPuncttrue
\mciteSetBstMidEndSepPunct{\mcitedefaultmidpunct}
{\mcitedefaultendpunct}{\mcitedefaultseppunct}\relax
\EndOfBibitem
\bibitem[Diederich and Kivala(2010)]{diederich2010}
Diederich,~F.; Kivala,~M. \emph{Adv. Mater.} \textbf{2010}, \emph{22},
  803--812\relax
\mciteBstWouldAddEndPuncttrue
\mciteSetBstMidEndSepPunct{\mcitedefaultmidpunct}
{\mcitedefaultendpunct}{\mcitedefaultseppunct}\relax
\EndOfBibitem
\bibitem[Haley(2008)]{haley2008}
Haley,~M.~M. \emph{Pure Appl. Chem.} \textbf{2008}, \emph{80}, 519--532\relax
\mciteBstWouldAddEndPuncttrue
\mciteSetBstMidEndSepPunct{\mcitedefaultmidpunct}
{\mcitedefaultendpunct}{\mcitedefaultseppunct}\relax
\EndOfBibitem
\bibitem[Kehoe et~al.(2000)Kehoe, Kiley, English, Johnson, Petersen, and
  Haley]{kehoe2000}
Kehoe,~J.~M.; Kiley,~J.~H.; English,~J.~J.; Johnson,~C.~A.; Petersen,~R.~C.;
  Haley,~M.~M. \emph{Org. Lett.} \textbf{2000}, \emph{2}, 969--972\relax
\mciteBstWouldAddEndPuncttrue
\mciteSetBstMidEndSepPunct{\mcitedefaultmidpunct}
{\mcitedefaultendpunct}{\mcitedefaultseppunct}\relax
\EndOfBibitem
\bibitem[Bunz et~al.(1999)Bunz, Rubin, and Tobe]{bunz1999}
Bunz,~U. H.~F.; Rubin,~Y.; Tobe,~Y. \emph{Chem. Soc. Rev.} \textbf{1999},
  \emph{28}, 107--119\relax
\mciteBstWouldAddEndPuncttrue
\mciteSetBstMidEndSepPunct{\mcitedefaultmidpunct}
{\mcitedefaultendpunct}{\mcitedefaultseppunct}\relax
\EndOfBibitem
\bibitem[Liu et~al.(2012)Liu, Xu, Li, and Li]{liu2012}
Liu,~H.; Xu,~J.; Li,~Y.; Li,~Y. \emph{Acc. Chem. Res.} \textbf{2012},
  \emph{43}, 1496--1508\relax
\mciteBstWouldAddEndPuncttrue
\mciteSetBstMidEndSepPunct{\mcitedefaultmidpunct}
{\mcitedefaultendpunct}{\mcitedefaultseppunct}\relax
\EndOfBibitem
\bibitem[Topsakal et~al.(2009)Topsakal, Akturk, and Ciraci]{bnhoneycomb}
Topsakal,~M.; Akturk,~E.; Ciraci,~S. \emph{Phys. Rev. B} \textbf{2009},
  \emph{79}, 115442\relax
\mciteBstWouldAddEndPuncttrue
\mciteSetBstMidEndSepPunct{\mcitedefaultmidpunct}
{\mcitedefaultendpunct}{\mcitedefaultseppunct}\relax
\EndOfBibitem
\bibitem[Narita et~al.(1998)Narita, Nagai, Suzuki, and Nakao]{narita1998}
Narita,~N.; Nagai,~S.; Suzuki,~S.; Nakao,~K. \emph{Phys. Rev. B} \textbf{1998},
  \emph{58}, 11009--11014\relax
\mciteBstWouldAddEndPuncttrue
\mciteSetBstMidEndSepPunct{\mcitedefaultmidpunct}
{\mcitedefaultendpunct}{\mcitedefaultseppunct}\relax
\EndOfBibitem
\bibitem[Coluci et~al.(2003)Coluci, Braga, Legoas, Galvao, and
  Baughman]{coluci2003}
Coluci,~V.~R.; Braga,~S.~F.; Legoas,~S.~B.; Galvao,~D.~S.; Baughman,~R.~H.
  \emph{Phys. Rev. B} \textbf{2003}, \emph{68}, 035430\relax
\mciteBstWouldAddEndPuncttrue
\mciteSetBstMidEndSepPunct{\mcitedefaultmidpunct}
{\mcitedefaultendpunct}{\mcitedefaultseppunct}\relax
\EndOfBibitem
\bibitem[Zhou et~al.(2011)Zhou, Lv, Wang, Chen, Sun, and Jena]{zhou2011}
Zhou,~J.; Lv,~K.; Wang,~Q.; Chen,~X.~S.; Sun,~Q.; Jena,~P. \emph{J. Chem.
  Phys.} \textbf{2011}, \emph{134}, 174701\relax
\mciteBstWouldAddEndPuncttrue
\mciteSetBstMidEndSepPunct{\mcitedefaultmidpunct}
{\mcitedefaultendpunct}{\mcitedefaultseppunct}\relax
\EndOfBibitem
\bibitem[Pan et~al.(2011)Pan, Zhang, Song, Du, and Gao]{pan2011}
Pan,~L.~D.; Zhang,~L.~Z.; Song,~B.~Q.; Du,~S.~X.; Gao,~H.~J. \emph{Appl. Phys.
  Lett.} \textbf{2011}, \emph{98}, 173102\relax
\mciteBstWouldAddEndPuncttrue
\mciteSetBstMidEndSepPunct{\mcitedefaultmidpunct}
{\mcitedefaultendpunct}{\mcitedefaultseppunct}\relax
\EndOfBibitem
\bibitem[Malko et~al.(2012)Malko, Neiss, Vines, and Gorling]{malko_hetero}
Malko,~D.; Neiss,~C.; Vines,~F.; Gorling,~A. \emph{Phys. Rev. B} \textbf{2012},
  \emph{86}, 045443\relax
\mciteBstWouldAddEndPuncttrue
\mciteSetBstMidEndSepPunct{\mcitedefaultmidpunct}
{\mcitedefaultendpunct}{\mcitedefaultseppunct}\relax
\EndOfBibitem
\bibitem[Sahin et~al.(2009)Sahin, Ataca, and Ciraci]{graphane}
Sahin,~H.; Ataca,~C.; Ciraci,~S. \emph{Appl. Phys. Lett.} \textbf{2009},
  \emph{95}, 222510\relax
\mciteBstWouldAddEndPuncttrue
\mciteSetBstMidEndSepPunct{\mcitedefaultmidpunct}
{\mcitedefaultendpunct}{\mcitedefaultseppunct}\relax
\EndOfBibitem
\bibitem[Blochl(1994)]{blochl1994}
Blochl,~P.~E. \emph{Phys. Rev. B} \textbf{1994}, \emph{50}, 17953--17979\relax
\mciteBstWouldAddEndPuncttrue
\mciteSetBstMidEndSepPunct{\mcitedefaultmidpunct}
{\mcitedefaultendpunct}{\mcitedefaultseppunct}\relax
\EndOfBibitem
\bibitem[Kresse and Hafner(1993)]{kresse1993}
Kresse,~G.; Hafner,~J. \emph{Phys. Rev. B} \textbf{1993}, \emph{47},
  558--561\relax
\mciteBstWouldAddEndPuncttrue
\mciteSetBstMidEndSepPunct{\mcitedefaultmidpunct}
{\mcitedefaultendpunct}{\mcitedefaultseppunct}\relax
\EndOfBibitem
\bibitem[Kresse and Furthmuller(1996)]{kresse1996}
Kresse,~G.; Furthmuller,~J. \emph{Phys. Rev. B} \textbf{1996}, \emph{54},
  11169--11186\relax
\mciteBstWouldAddEndPuncttrue
\mciteSetBstMidEndSepPunct{\mcitedefaultmidpunct}
{\mcitedefaultendpunct}{\mcitedefaultseppunct}\relax
\EndOfBibitem
\bibitem[Perdew et~al.(1992)Perdew, Chevary, Vosko, Jackson, Pederson, Singh,
  and Fiolhais]{perdew1992}
Perdew,~J.~P.; Chevary,~J.~A.; Vosko,~S.~H.; Jackson,~K.~A.; Pederson,~M.~R.;
  Singh,~D.~J.; Fiolhais,~C. \emph{Phys. Rev. B} \textbf{1992}, \emph{46},
  6671--6687\relax
\mciteBstWouldAddEndPuncttrue
\mciteSetBstMidEndSepPunct{\mcitedefaultmidpunct}
{\mcitedefaultendpunct}{\mcitedefaultseppunct}\relax
\EndOfBibitem
\bibitem[Grimme(2006)]{grimme2006}
Grimme,~S. \emph{J. Comput. Chem.} \textbf{2006}, \emph{27}, 1787--1799\relax
\mciteBstWouldAddEndPuncttrue
\mciteSetBstMidEndSepPunct{\mcitedefaultmidpunct}
{\mcitedefaultendpunct}{\mcitedefaultseppunct}\relax
\EndOfBibitem
\bibitem[Monkhorst and Pack(1976)]{monkhorst1976}
Monkhorst,~H.~J.; Pack,~J.~D. \emph{J. Comput. Chem.} \textbf{1976}, \emph{13},
  5188\relax
\mciteBstWouldAddEndPuncttrue
\mciteSetBstMidEndSepPunct{\mcitedefaultmidpunct}
{\mcitedefaultendpunct}{\mcitedefaultseppunct}\relax
\EndOfBibitem
\bibitem[Giannozzi(2009)]{giannozzi2009}
Giannozzi,~P. e.~a. \emph{J. Phys.: Condens. Matter} \textbf{2009}, \emph{21},
  395502\relax
\mciteBstWouldAddEndPuncttrue
\mciteSetBstMidEndSepPunct{\mcitedefaultmidpunct}
{\mcitedefaultendpunct}{\mcitedefaultseppunct}\relax
\EndOfBibitem
\bibitem[Tongay et~al.(2004)Tongay, Senger, Dag, and Ciraci]{carbonchain1}
Tongay,~S.; Senger,~R.~T.; Dag,~S.; Ciraci,~S. \emph{Phys. Rev. Lett.}
  \textbf{2004}, \emph{93}, 136404\relax
\mciteBstWouldAddEndPuncttrue
\mciteSetBstMidEndSepPunct{\mcitedefaultmidpunct}
{\mcitedefaultendpunct}{\mcitedefaultseppunct}\relax
\EndOfBibitem
\bibitem[Pauling(1960)]{pauling}
Pauling,~L. \emph{The Nature of the Chemical Bonds, 3rd Edition};
\newblock Cornell University Press: New York, 1960\relax
\mciteBstWouldAddEndPuncttrue
\mciteSetBstMidEndSepPunct{\mcitedefaultmidpunct}
{\mcitedefaultendpunct}{\mcitedefaultseppunct}\relax
\EndOfBibitem
\bibitem[Eisler et~al.(2005)Eisler, Aaron, Elliott, Luu, McDonald, Hegmann, and
  Tykwinski]{eisler}
Eisler,~S.; Aaron,~D.; Elliott,~E.; Luu,~T.; McDonald,~R.; Hegmann,~F.~A.;
  Tykwinski,~R.~R. \emph{J. Am. Chem. Soc.} \textbf{2005}, \emph{127},
  2666--26676\relax
\mciteBstWouldAddEndPuncttrue
\mciteSetBstMidEndSepPunct{\mcitedefaultmidpunct}
{\mcitedefaultendpunct}{\mcitedefaultseppunct}\relax
\EndOfBibitem
\bibitem[Meyer et~al.(2008)Meyer, Girit, Crommmie, and Zettl]{meyer}
Meyer,~J.~C.; Girit,~C.~O.; Crommmie,~M.~F.; Zettl,~A. \emph{Nature (London)}
  \textbf{2008}, \emph{454}, 319--322\relax
\mciteBstWouldAddEndPuncttrue
\mciteSetBstMidEndSepPunct{\mcitedefaultmidpunct}
{\mcitedefaultendpunct}{\mcitedefaultseppunct}\relax
\EndOfBibitem
\bibitem[Chalifoux and Tykwinski(2010)]{chalifoux}
Chalifoux,~W.~A.; Tykwinski,~R.~R. \emph{Nature Chem.} \textbf{2010}, \emph{2},
  967--971\relax
\mciteBstWouldAddEndPuncttrue
\mciteSetBstMidEndSepPunct{\mcitedefaultmidpunct}
{\mcitedefaultendpunct}{\mcitedefaultseppunct}\relax
\EndOfBibitem
\bibitem[Kittel(1996)]{kittel}
Kittel,~C. \emph{Introduction to Solid State Physics, 8th Edition};
\newblock John Wiley \& Sons: New York, 1996\relax
\mciteBstWouldAddEndPuncttrue
\mciteSetBstMidEndSepPunct{\mcitedefaultmidpunct}
{\mcitedefaultendpunct}{\mcitedefaultseppunct}\relax
\EndOfBibitem
\bibitem[pol()]{poly}
 It should be noted that the infinite carbon chain can undergo a Peierls
  distortion; the energy is slightly lowered and alternating carbon atoms are
  slightly displaced from their equilibrium positions in cumulene. At the end,
  the unit cell consisting of two atoms bound by short (triple) and long
  (single) bonds is doubled. This structure is called polyne.\relax
\mciteBstWouldAddEndPunctfalse
\mciteSetBstMidEndSepPunct{\mcitedefaultmidpunct}
{}{\mcitedefaultseppunct}\relax
\EndOfBibitem
\bibitem[Tongay et~al.(2004)Tongay, Durgun, and Ciraci]{bnchain}
Tongay,~S.; Durgun,~E.; Ciraci,~S. \emph{Appl. Phys. Lett.} \textbf{2004},
  \emph{85}, 6179\relax
\mciteBstWouldAddEndPuncttrue
\mciteSetBstMidEndSepPunct{\mcitedefaultmidpunct}
{\mcitedefaultendpunct}{\mcitedefaultseppunct}\relax
\EndOfBibitem
\bibitem[Cahangirov et~al.(2010)Cahangirov, Topsakal, and Ciraci]{carbonchain}
Cahangirov,~S.; Topsakal,~M.; Ciraci,~S. \emph{Phys. Rev. B} \textbf{2010},
  \emph{82}, 195444\relax
\mciteBstWouldAddEndPuncttrue
\mciteSetBstMidEndSepPunct{\mcitedefaultmidpunct}
{\mcitedefaultendpunct}{\mcitedefaultseppunct}\relax
\EndOfBibitem
\bibitem[Cahangirov et~al.(2009)Cahangirov, Topsakal, Akturk, Sahin, and
  Ciraci]{silicene}
Cahangirov,~S.; Topsakal,~M.; Akturk,~E.; Sahin,~H.; Ciraci,~S. \emph{Phys.
  Rev. Lett.} \textbf{2009}, \emph{102}, 236804\relax
\mciteBstWouldAddEndPuncttrue
\mciteSetBstMidEndSepPunct{\mcitedefaultmidpunct}
{\mcitedefaultendpunct}{\mcitedefaultseppunct}\relax
\EndOfBibitem
\bibitem[Ataca and Ciraci(2011)]{can_chain}
Ataca,~C.; Ciraci,~S. \emph{Phys. Rev. B} \textbf{2011}, \emph{83},
  235417\relax
\mciteBstWouldAddEndPuncttrue
\mciteSetBstMidEndSepPunct{\mcitedefaultmidpunct}
{\mcitedefaultendpunct}{\mcitedefaultseppunct}\relax
\EndOfBibitem
\bibitem[\"{O}z\c celik and Ciraci(2012)]{ongun_chain}
\"{O}z\c celik,~V.~O.; Ciraci,~S. \emph{Phys. Rev. B} \textbf{2012}, \emph{86},
  155421\relax
\mciteBstWouldAddEndPuncttrue
\mciteSetBstMidEndSepPunct{\mcitedefaultmidpunct}
{\mcitedefaultendpunct}{\mcitedefaultseppunct}\relax
\EndOfBibitem
\bibitem[\"{O}z\c celik et~al.(2012)\"{O}z\c celik, Cahangirov, and
  Ciraci]{growth}
\"{O}z\c celik,~V.~O.; Cahangirov,~S.; Ciraci,~S. \emph{Phys. Rev. B}
  \textbf{2012}, \emph{85}, 235456\relax
\mciteBstWouldAddEndPuncttrue
\mciteSetBstMidEndSepPunct{\mcitedefaultmidpunct}
{\mcitedefaultendpunct}{\mcitedefaultseppunct}\relax
\EndOfBibitem
\bibitem[Topsakal et~al.(2010)Topsakal, Cahangirov, and Ciraci]{topsakal2010}
Topsakal,~M.; Cahangirov,~S.; Ciraci,~S. \emph{Appl. Phys. Lett.}
  \textbf{2010}, \emph{96}, 091912\relax
\mciteBstWouldAddEndPuncttrue
\mciteSetBstMidEndSepPunct{\mcitedefaultmidpunct}
{\mcitedefaultendpunct}{\mcitedefaultseppunct}\relax
\EndOfBibitem
\bibitem[Cranford and Buehler(2011)]{cranford2011}
Cranford,~S.~W.; Buehler,~M.~J. \emph{Carbon} \textbf{2011}, \emph{49},
  4111--4121\relax
\mciteBstWouldAddEndPuncttrue
\mciteSetBstMidEndSepPunct{\mcitedefaultmidpunct}
{\mcitedefaultendpunct}{\mcitedefaultseppunct}\relax
\EndOfBibitem
\bibitem[Kang et~al.(2011)Kang, Li, Wu, Li, and Xia]{kang2011}
Kang,~J.; Li,~J.; Wu,~F.; Li,~S.~S.; Xia,~J.~B. \emph{J. Phys. Chem. C}
  \textbf{2011}, \emph{115}, 20466--20470\relax
\mciteBstWouldAddEndPuncttrue
\mciteSetBstMidEndSepPunct{\mcitedefaultmidpunct}
{\mcitedefaultendpunct}{\mcitedefaultseppunct}\relax
\EndOfBibitem
\bibitem[Lee et~al.(2008)Lee, Wei, Kysar, and Hone]{lee2008}
Lee,~C.; Wei,~X.; Kysar,~J.~W.; Hone,~J. \emph{Science} \textbf{2008},
  \emph{321}, 385--388\relax
\mciteBstWouldAddEndPuncttrue
\mciteSetBstMidEndSepPunct{\mcitedefaultmidpunct}
{\mcitedefaultendpunct}{\mcitedefaultseppunct}\relax
\EndOfBibitem
\bibitem[Sahin and Ciraci(2011)]{sahin2011}
Sahin,~H.; Ciraci,~S. \emph{Phys. Rev. B} \textbf{2011}, \emph{84},
  035452\relax
\mciteBstWouldAddEndPuncttrue
\mciteSetBstMidEndSepPunct{\mcitedefaultmidpunct}
{\mcitedefaultendpunct}{\mcitedefaultseppunct}\relax
\EndOfBibitem
\bibitem[Sofo et~al.(2007)Sofo, Chaudhari, and Barber]{sofo2007}
Sofo,~J.~O.; Chaudhari,~A.~S.; Barber,~G.~D. \emph{Phys. Rev. B} \textbf{2007},
  \emph{75}, 153401\relax
\mciteBstWouldAddEndPuncttrue
\mciteSetBstMidEndSepPunct{\mcitedefaultmidpunct}
{\mcitedefaultendpunct}{\mcitedefaultseppunct}\relax
\EndOfBibitem
\bibitem[Averill et~al.(2009)Averill, Morris, and Cooper]{averill2009}
Averill,~F.~W.; Morris,~J.~R.; Cooper,~V.~R. \emph{Phys. Rev. B} \textbf{2009},
  \emph{80}, 195411\relax
\mciteBstWouldAddEndPuncttrue
\mciteSetBstMidEndSepPunct{\mcitedefaultmidpunct}
{\mcitedefaultendpunct}{\mcitedefaultseppunct}\relax
\EndOfBibitem
\bibitem[Zheng et~al.(2012)Zheng, Luo, Liu, Quhe, Zheng, Tang, Gao, Nagase, and
  Lu]{zheng2012}
Zheng,~Q.; Luo,~G.; Liu,~Q.; Quhe,~R.; Zheng,~J.; Tang,~K.; Gao,~Z.;
  Nagase,~S.; Lu,~J. \emph{Nanoscale} \textbf{2012}, \emph{4}, 3990--3996\relax
\mciteBstWouldAddEndPuncttrue
\mciteSetBstMidEndSepPunct{\mcitedefaultmidpunct}
{\mcitedefaultendpunct}{\mcitedefaultseppunct}\relax
\EndOfBibitem
\bibitem[Bj\"orkman et~al.(2012)Bj\"orkman, Gulans, Krasheninnikov, and
  Nieminen]{bjorkman2012}
Bj\"orkman,~T.; Gulans,~A.; Krasheninnikov,~A.~V.; Nieminen,~R.~M. \emph{Phys.
  Rev. Lett.} \textbf{2012}, \emph{108}, 235501\relax
\mciteBstWouldAddEndPuncttrue
\mciteSetBstMidEndSepPunct{\mcitedefaultmidpunct}
{\mcitedefaultendpunct}{\mcitedefaultseppunct}\relax
\EndOfBibitem
\bibitem[Partoens and Peeters(2006)]{partoens2006}
Partoens,~B.; Peeters,~F.~M. \emph{Phys. Rev. B} \textbf{2006}, \emph{74},
  075404\relax
\mciteBstWouldAddEndPuncttrue
\mciteSetBstMidEndSepPunct{\mcitedefaultmidpunct}
{\mcitedefaultendpunct}{\mcitedefaultseppunct}\relax
\EndOfBibitem
\bibitem[Durgun et~al.(2006)Durgun, Senger, Mehrez, Dag, and Ciraci]{engin2006}
Durgun,~E.; Senger,~R.~T.; Mehrez,~H.; Dag,~S.; Ciraci,~S. \emph{Europhys.
  Lett.} \textbf{2006}, \emph{73}, 642--648\relax
\mciteBstWouldAddEndPuncttrue
\mciteSetBstMidEndSepPunct{\mcitedefaultmidpunct}
{\mcitedefaultendpunct}{\mcitedefaultseppunct}\relax
\EndOfBibitem
\end{mcitethebibliography}

\end{document}